\documentclass[prb,twocolumn,superscriptaddress,floatfix]{revtex4-1}
\usepackage[english]{babel}
\usepackage{graphicx}
\usepackage{amsmath}
\usepackage{amssymb}
\begin{document}
\title{Spin relaxation in CdTe quantum dots with a single Mn atom}
\author{Marko D. Petrovi\'c} 
\affiliation{Scientific Computing Laboratory, Institute of Physics Belgrade, University of Belgrade, 
Pregrevica 118, 11080 Belgrade, Serbia}
\author{Nenad Vukmirovi\'c}\email{nenad.vukmirovic@ipb.ac.rs}
\affiliation{Scientific Computing Laboratory, Institute of Physics Belgrade, University of Belgrade, 
Pregrevica 118, 11080 Belgrade, Serbia}

\begin{abstract} 
We have investigated spin relaxation times in CdTe quantum dots doped with a single
Mn atom, a prototype of a system where the interaction between a single charge carrier and a single spin takes place. A theoretical model that was used includes the electron--Mn spin exchange interaction responsible for mixing of the states of different spin in the basic Hamiltonian and electron--phonon interaction as a perturbation responsible for transitions between the states.
It was found that the dominant electron--phonon interaction mechanism responsible for spin relaxation is the interaction with acoustic phonons through deformation potential. Electron and Mn spin relaxation times at room temperature take values in the range from microseconds at a magnetic field of 0.5~T down to nanoseconds at a magnetic field of 10~T and become three orders of magnitude larger at cryogenic temperatures. It was found that electron spin-orbit interaction has a negligible effect on spin relaxation times, while the changes in the position of the Mn atom within the dot and in the dot dimensions can change the spin relaxation times by up to one order of magnitude. 
%\textbf{I ovaj komentar treba dobro proveriti.} Our results provide useful information on inherent limitations to the control of spin states in similar systems and suggest some routes for overcoming them. \textbf{Ovu poslednju recenicu ostaviti samo ako to budem elaborirao detaljnije u sekciji 4- diskusija ili zakljucku.}

\end{abstract} 

%mesta za citiranje:
%mesta za citiranje:
%1. nesto da su vreme zivota spina bitna za QIP i Spintroniku - stavio sam
%2. eksperimenti o kontroli Mn spina i teorijske predikcije za istu - stavio sam
%3. lukas teorijska analiza - komentarisano
%4. najnoviji rad o Mn - phonon coupling-u - stavio sam
%6. raniji radovi o relaksaciji spina u kvantnim tackama bez Mn atoma - stavio sam
%8. drugi mehanizmi relaksacije
%     -Rashba spin-orbit coupling - komentarisano
%     -hyperfine interaction with nuclei - komentarisano
%     -direktan elektron spin-phonon coupling - komentarisano
%     -direct Mn-spin coupling - komentarisano
%     -....
%9. reference za parametre materijala
%11. komentari rezultata iz ranijih radova
%   -uticaj dimenzija (oscilacije)
%   -uticaj magnetskog polja (oscilacije)
%   -utacija temperature - valjda je svuda isto.
%   -mozda neki komentar o spin-orbit interakciji.
%   -generalno o vremenima zivota (i kolika bi bila da utice samo SO interakcija).
%%%%%%%%%%%%%%%%%%%%%%%%%%%%%%%%%%%%%%%%%%%%55
%10. ranije studije energetskog spektra u Mn kvantnim tackama.
%5. raniji radovi o Mn kvantnim tackama, ali bez analize vremena zivota spina
%7. Hamiltonijan iz ranijih radova
%Objasniti jos u vezi usrednjavanja na slikama.

\maketitle{}

\section{Introduction}

The potential of utilizing the spin degree of freedom as a classical bit in 
spintronic devices\cite{rmp76-323,pr493-61,prb84-024402} and as a quantum bit in
potential quantum information processing devices\cite{Awschalom} has been largely recognized
in the last two decades. Operation of these devices crucially depends on the 
ability to manipulate the spin degree of freedom of the system.\cite{HeuleEtAl} However, 
in all realistic open systems, undesirable spin flips occur due to the interaction
with environment. It is therefore of great importance to understand and be able to 
quantitatively describe the mechanisms of spin dephasing and spin relaxation. 

Physical systems which are expected to be particularly suitable for aforementioned 
applications are based on architectures that contain quantum dots -- artificial nanostructures where 
charge carriers are confined in all three spatial directions.\cite{Bimberg,Harrison} Due to quantum confinement
effect the spectrum of electronic states in quantum dots is discrete and as a consequence
phase space for relaxation and dephasing processes is greatly reduced.\cite{prb42-8947} Therefore, 
long spin lifetimes of electrons confined in quantum dots are expected. 

In addition, quantum dots provide a playground where one can study fundamental interactions
on a single carrier or spin level. Such a level of understanding is necessary before one 
can proceed to understand more complicated device structures. Quantum dots doped with a 
single Mn atom have drawn particular attention for fundamental studies in recent years. The 
manganese atom acts effectively as an additional spin 5/2 degree of freedom and therefore enables
fundamental studies of interaction between a single charge and a single spin in these dots. 
Experimentally observed signature of this interaction is the splitting of an exciton line in quantum dot 
photoluminescence spectrum.\cite{prl93-207403,prl95-047403,prb71-161307,prb73-045301,prb68-045303}
Predictions that it is possible to optically manipulate the state of a Mn spin in the 
quantum dot~\cite{prb71-035338,prb73-045301,prl102-177403} were recently realized in several experiments.~\cite{prl102-127402,prb81-245315,prl103-087401}

However, very little is known about the lifetimes of Mn and electron spin in these dots. 
A theoretical analysis\cite{prb82-075321} of optical orientation experiments\cite{prb81-245315} 
provided estimates of hole and electron spin relaxation times necessary for Mn spin orientation to occur. 
In a very recent work, Mn spin relaxation times in quantum dots
in the presence of a hole or an exciton were calculated based on a microscopic theory.\cite{prb84-205305}
Other theoretical studies of quantum dots with Mn 
atoms\cite{prb76-245307,crp9-857,prl95-217206,prb74-245308,prl96-157201,prb72-075358,prb72-075359,prl95-117201,prl98-106805,prb78-045321,prb76-045315}  were focused on
electronic, optical, magnetic or transport properties without any discussion of spin lifetimes. Numerous studies of spin relaxation in quantum dots were 
restricted to quantum dots without a Mn atom~\cite{prb61-12639,prb66-155327,prb66-161318,prb67-205330,prb69-115318,prb69-125330,prb71-075308,prb71-205324,prb72-115326,prb72-155333,prb75-041306}
or on diluted magnetic semiconductor quantum dots with many Mn atoms.\cite{prb72-075303}

In this work, we calculate the relaxation times of Mn and electron
spin caused by interaction with phonons in a singly negatively charged CdTe quantum dot containing one Mn atom. 
Our calculation is based on a microscopic theory that links the system geometry with spin relaxation times. 
In Sec.~\ref{Sec:theory} we introduce the theoretical model used to describe the system at hand and 
relevant spin relaxation times. In Sec.~\ref{Sec:results} we present the results and analyze the 
effects of different electron--phonon interaction mechanisms, spin-orbit (SO) interaction, magnetic field, 
temperature, quantum dot dimensions and Mn atom position. In Sec.~\ref{Sec:discus} we compare our
results to theoretical results for spin relaxation in somewhat similar systems and experiments,
and analyze the strength of other possible spin relaxation mechanisms not included in our model.

\section{Theoretical model}\label{Sec:theory}  
In this section, we describe a theoretical model used to describe a CdTe quantum dot that contains one 
extra electron and one Mn atom placed in its interior. Such a scenario is experimentally realized in 
a layer of self-assembled quantum dots grown at relatively low dot density such that interdot 
interactions are negligible.
                           
The Hamiltonian of the system reads                                           
     \begin{equation}                                                        
     \hat{H}_{o}=\hat{H}_{el}+\hat{H}_{m\textrm{-}el}+\hat{H}_B,             
     \end{equation}                                                          
where $\hat{H}_{el}$ is the electronic Hamiltonian, $\hat{H}_{m\textrm{-}el}$ describes the interaction
of electron with Mn spin and $\hat{H}_B$ is the Zeeman term that describes the interaction of electron 
and Mn spin with external magnetic field. This Hamiltonian acts in the Hilbert space of the system 
which is given as the direct product of electron orbital space, electron spin space and Mn spin space.

The first, electronic term reads                                    
    \begin{equation}                                                         
    \hat{H}_{el}=\sum_{ i, \sigma}E_{i,\sigma}\hat{c}^\dagger_{i,\sigma}     
                 \hat{c}_{i,\sigma},                                         
    \end{equation}                                                           
with $\hat{c}^\dagger_{i,\sigma}$ and $\hat{c}_{i,\sigma}$ representing 
electron creation and annihilation operators. Typical self-assembled quantum dots are much larger in 
the lateral plane ($xy$ plane) than in the growth ($z$) direction, while they can take various shapes --
lenses, pyramids, truncated pyramids, etc. For this reason, we adopt the simplest possible quantum dot
model that captures all essential features of the single-particle electronic spectrum -- a rectangular
 box with in-plane dimensions ($L_x$ and $L_y$) much larger than the dimension in the $z$ direction ($L_z$)
with infinite potential barriers outside of the region $0\le x\le L_x$, $0\le y\le L_y$, $0\le z\le L_z$.
The electronic states in the conduction band of semiconductor nanostructures within first several 
hundreds of meV are well described with envelope function effective mass Hamiltonian. For our quantum dot
 model, the conduction band electron envelope function is then given as
    \begin{equation}                                                         \label{eq:psieis}
    \psi(x,y,z) = N_c
\sin\left(\frac{n_x\pi}{L_x}x\right)
\sin\left(\frac{n_y\pi}{L_y}y\right)        
\sin\left(\frac{n_z\pi}{L_z}z\right)                                  
    \end{equation}                                                           
where $N_c=\sqrt{{8}/\left(L_xL_yL_z\right)}$ is the normalization constant and $n_x$, 
$n_y$ and $n_z$ are positive integers that represent orbital quantum numbers in each direction.
Single particle energy $E_{i,\sigma}$ of an electronic state $i$ with orbital quantum numbers $n_x$, 
$n_y$ and $n_z$ is given as
\begin{equation}\label{eq:eis}
E_{i,\sigma}=\frac{\hbar^2\pi^2}{2m^*}\left(\frac{n_x^2}{L_x^2}+\frac{n_y^2}{L_y^2}+\frac{n_z^2}{L_z^2}\right),
\end{equation}
where $m^*$ is the effective mass of the conduction band electron in CdTe.
                                                                           
The second term $\hat{H}_{m\textrm{-}el}$ describes the exchange interaction between an
 electron and a Mn atom, and it is given by the spin impurity model Hamiltonian,\cite{Mahan} a 
model well established in previous theoretical studies of CdTe quantum dots with a few Mn
atoms\cite{prl95-217206,prl96-157201,prb74-245308,prb76-245307,prb78-045321,prb76-045315}                                    
    \begin{eqnarray} \label{eq:jijr}                  
    \hat{H}_{m\textrm{-}el} = -\frac{1}{2} \sum_{i,j} J_{ij}(\mathbf{R})      
           \left[                                                            
                (\hat{c}^\dagger_{i,\uparrow} \hat{c}_{j,\uparrow} -         
                 \hat{c}^\dagger_{i,\downarrow} \hat{c}_{j,\downarrow})      
                 \hat{M}_{z}                                                 
            \right.                                       \nonumber \\       
          \left.                                                             
               + \hat{c}^\dagger_{i,\downarrow} \hat{c}_{j,\uparrow}         
               \hat{M}^{+}                                                   
               + \hat{c}^\dagger_{i,\uparrow} \hat{c}_{j,\downarrow}         
               \hat{M}^{-}                                                   
          \right].                                                           
    \end{eqnarray}                                                           
%
%     stao2                                                                                                                            
In Eq.~(ref{eq:jijr}), $\mathbf{R}$ is the position of the Mn atom and $J_{ij}
(\mathbf{R})$ is the electron--Mn spin coupling strength, equal to $J_{c}\psi^
{*}_{i}(\mathbf{R})\psi_{j}(\mathbf{R})$. $\hat{M}_{z}$, $\hat{M}^{+}$ and 
$\hat{M}^{-}$ are the Mn spin operators whose properties are governed by the 
spin 5/2 algebra.

The last term of $\hat{H}_{o}$, also known as the Zeeman term, 
describes the interaction of the whole system with an external magnetic 
field $B$ parallel to the $z$ direction. It is given as                                                   
    \begin{equation}  \label{eq:muBB}                                                       
    \hat{H}_{B} = -\mu_{B} g_{e} B \hat{S}_z -\mu_B g_{Mn} B \hat{M}_z,      
    \end{equation}                                                           
where $\mu_{B}$ is the Bohr magneton, and $g_{e}$ ($g_{Mn}$) is the electron spin
(Mn spin) $g$ factor. $\hat{S}_{z}$ is the operator of the $z$ component of the 
electron spin.

In Sec.~\ref{Sec:soint}, we also consider electronic SO coupling
that arises in materials lacking inversion symmetry (Dresselhaus SO coupling\cite{pr100-580})                                                
    \begin{equation}         \label{eq:hso}                                                
    \hat{H}_{so} = \gamma \mathbf{h} \cdot \mathbf{\sigma},                        
    \end{equation}                                                           
where $\mathbf{\sigma}$-s are Pauli matrices, $\mathbf{h}$ is the Dresselhaus effective magnetic field 
    \begin{equation}                                                         
    \mathbf{h}=[k_x(k_y^2-k_z^2),k_y(k_z^2-k_x^2),k_z(k_x^2-k_y^2)],         
    \end{equation}                                                           
with $\mathbf{k}=-i \: \mathbf{\nabla}$ and $\gamma$ is the Dresselhaus SO coupling strength.

After numerically solving the Hamiltonian eigenvalue problem, we obtain the eigenenergies $E_{a}$ and eigenstates $|\Psi_{a}\rangle$
    \begin{eqnarray}                     \label{eq:psia}                                    
    \hat{H}_{o}|\Psi_{a}\rangle & = & E_{a}|\Psi_{a}\rangle,\nonumber\\               
    |\Psi_{a}\rangle & = & \sum_{ i, S_{z}, M_{z}} c^{a}_{i S_z M_z}         
                           | i, S_{z}, M_{z} \rangle.                        
    \end{eqnarray}                                                           
We use $i$ and $S_z$ to denote the orbital and spin state of the electron and $M_z$ for 
the Mn spin state.

Electron--phonon interaction is considered to be the main mechanism responsible
for the transitions between the eigenstates $|\Psi_{a}\rangle$ and consequently spin relaxation. 
We will show in Sec.~\ref{Sec:energy_spectrum} that relevant transition energies are of the order of meV,
which are typical energies of acoustic phonons. Because of the high energy of optical  phonons
compared to the acoustic ones, we consider only the interaction with acoustic phonons.
The Hamiltonian of electron--phonon interaction is given as\cite{Mahan,Harrison}                                                                      
    \begin{equation}                                                         
    \hat{H}_{e\textrm{-}ph} =                                                
              \sum_{ \mathbf{q}, \lambda} M_{ \mathbf{q}, \lambda}           
              ( \hat{b}^{ \dagger}_{ \mathbf{q}, \lambda} +                  
                \hat{b}_{ -\mathbf{q}, \lambda})                             
              e^{ i \mathbf{q} \mathbf{r} },                                               
    \end{equation}                                                           
where $\hat{b}^{\dagger}$ and $\hat{b}$ are phonon creation and annihilation 
operators and $\mathbf{q}$ is the phonon wave vector. For acoustic phonons, 
a linear dispersion relation connects the phonon wave vector and its
energy $\omega=qv$, where $v$ is the sound velocity for a particular 
acoustic phonon branch in a given material.

The scattering matrix $M_{\mathbf{q},\lambda}$ depends on the type of electron--phonon 
interaction. For the interaction through deformation potential it is given as\cite{Mahan}                                          
    \begin{equation} \label{eq:sme1}                                                        
   \left| M_{\mathbf{q},\lambda} \right|^2=                                                 
           \frac{ \hbar D^2 |\mathbf{q}| } { 2V \rho v_{\textrm{LA}} },      
    \end{equation}                                                           
and for the interaction through piezoelectric field it can be represented as\cite{Mahan}      
    \begin{equation}                \label{eq:sme2}                                         
   \left| M_{\mathbf{q},\lambda} \right|^2 = \frac{\hbar\xi} { v_{\textrm{LA}} }                               
                             \frac{\left(3q_xq_yq_z\right)^2} {|\mathbf{q}|^7}              
    \end{equation}                                                           
for longitudinal phonons and                                                  
    \begin{equation}                              \label{eq:sme3}                           
   \left| M_{\mathbf{q},\lambda} \right|^2 = \frac{ \hbar\xi }{ v_{\textrm{TA}} }            
           \left[                                                            
                 \frac{ q_x^2q_y^2+q^2_yq^2_z+q^2_zq^2_x }{|\mathbf{q}|^5}   
                 - \frac{\left(3q_xq_yq_z\right)^2}{|\mathbf{q}|^7}                         
           \right]                                                           
    \end{equation}                                                           
for transversal ones. For transversal acoustic phonons (TA) there are two branches 
($\lambda$=2, 3), whereas for longitudinal acoustic (LA) phonons there is only 
one branch ($\lambda$=1). In the preceding expressions $D$ is the 
acoustic deformation potential, $\rho$ is the CdTe material mass density, $V$ is the 
volume of the system and $\xi$ is given as
   \begin{equation}                                                          
   \xi=\frac{32\pi^2e^2h_{14}^2}{\kappa V \rho},        \nonumber                  
   \end{equation}                                                            
where $\kappa$ is the static dielectric constant and $h_{14}$ is the 
piezoelectric constant. 

For the work presented here, we have used the following 
parameters: 
\mbox{$J_{c}=15\:\mathrm{eV}\mathrm{\AA}^3$}, 
\mbox{$m^*=0.106\:m_0$} 
(Ref.~\onlinecite{prl95-217206}), 
\mbox{$g_\mathrm{e}=-1.67$}, 
\mbox{$g_\mathrm{Mn}=2.02$} 
(Ref.~\onlinecite{prl96-157201}), 
\mbox{$\gamma=11.74\:\mathrm{eV}\mathrm{\AA}^3$} 
(Ref. \onlinecite{prb38-1806}), 
\mbox{$v_\mathrm{LA}=3083\:\mathrm{m}/\mathrm{s}$}, 
\mbox{$v_\mathrm{TA}=1847\:\mathrm{m}/\mathrm{s}$}, 
\mbox{$h_\mathrm{14}=3.94\times10^8\:\mathrm{V}/\mathrm{m}$} (Ref.~\onlinecite{prb72-075303}), 
\mbox{$D=5.1\:\mathrm{eV}$}, 
\mbox{$\rho=4.85\times10^3\:\mathrm{kg}/\mathrm{m}^3$}, 
\mbox{$\kappa=9.6$} 
(Ref.~\onlinecite{bornstein}). 
Unless otherwise stated, quantum dot dimensions were taken as 
\mbox{$L_x=150\:\mathrm{\AA}$}, 
\mbox{$L_y=140\:\mathrm{\AA}$}, 
\mbox{$L_z= 30\:\mathrm{\AA}$}. 
In our calculations, we consider first \mbox{$N=10$} electron orbitals, which gives the 
Hilbert space of \mbox{$N_\mathrm{max}=120$} basis vectors. This ensures that the calculated quantities have converged to their 
real values.

Because of the fact that the electron--acoustic phonon interaction term is much smaller
compared to the rest of the Hamiltonian, this term can be treated as a perturbation responsible 
for transitions between eigenstates $|\Psi_{a}\rangle$ of $\hat{H}_{o}$.  
If the system starts in an initial state $|\Psi_i\rangle$, it will make a transition to a final state 
$|\Psi_f\rangle$ due to electron--phonon scattering. The scattering rate for this process is determined by Fermi's 
golden rule 
    \begin{eqnarray} \label{Eq:eqgif}                                                        
    \Gamma_{if} = \frac{2\pi}{\hbar}  \sum_{\mathbf{q},\lambda}              
                  |M_{\mathbf{q},\lambda}|^2  |\langle\Psi_{f}|              
                   e^{i\mathbf{q}\mathbf{r}}  |\Psi_{i}\rangle|^2            
                  \quad\quad\qquad\nonumber\\                               
    \times\left(                                                                  
           \bar{n}_{\mathbf{q},\lambda} + \frac{1}{2} \pm \frac{1}{2}        
        \right)                                                                    
        \delta\left(                                                                    
          E_{f} - E_{i} \pm \hbar \omega_{\mathbf{q}, \lambda} 
         \right),                                                                  
    \end{eqnarray}                                                           
where $E_{i}$ and $E_{f}$ are the energies of the unperturbed system in the
initial and the final state, $\bar{n}_{\mathbf{q},\lambda}$ is the mean number
of phonons at a given temperature and $|\langle\Psi_{f}|              
                   e^{i\mathbf{q}\mathbf{r}}  |\Psi_{i}\rangle|$ is the form factor
for electron--phonon interaction. The ($+$) sign in Eq.~(\ref{Eq:eqgif}) corresponds 
to the process of phonon emission, while the ($-$) sign corresponds to phonon absorption.

We will show in Sec.~\ref{Sec:results} that most of the eigenstates of $\hat{H}_{o}$ have a 
well defined Mn and electron spin because one of the $c_{i S_z M_z}$ 
coefficients in Eq.~(\ref{eq:psia}) is typically significantly larger than the others. Nevertheless, 
the remaining coefficients, that correspond to basis states with other values of $S_z$ and $M_z$, 
are nonzero. For this reason, the transitions between the states with different values of electron (or Mn) 
spin are allowed despite the fact that the electron--phonon interaction Hamiltonian is spin--independent.

To obtain the average electron 
%(or Mn) 
spin relaxation time one has to consider all possible transitions
in the system where a particular change of spin occurs. For example, the mean relaxation time for electron
spin change from the initial spin $S_z=1/2$ to the final spin $S_z=-1/2$ is given as
    \begin{equation}                \label{eq:avetau}                                         
    \frac{1}{\tau} = \frac
      {\sum_{i} f_i \sum_{f} \Gamma_{if} }                      
      {\sum_{i} f_i},                        
    \end{equation}                                                           
where the sum over $i$ includes all possible initial states with $S_z=1/2$ and the sum over 
$f$ includes all possible final states with $S_z=-1/2$, with $f_i$ being the thermal 
weighting factor of state $i$ at a temperature $T$, given as                                                                            
    $f_i = \exp\left[ - {E_i}/\left({k_BT}\right) \right]$.                                  

\section{Results}          \label{Sec:results}                  
                                                                              
\subsection{Energy spectrum}\label{Sec:energy_spectrum}
                     
The energy spectrum of CdTe quantum dots with a single Mn atom has been 
studied in the past\cite{prb72-235332,prb76-245307,prb78-045321} and here we only review
the main features, with an emphasis on those that are relevant for our work.
%stao3

\begin{figure}[bt]                                                       
    \centering                                                               
    \includegraphics[width= 0.49 \textwidth]{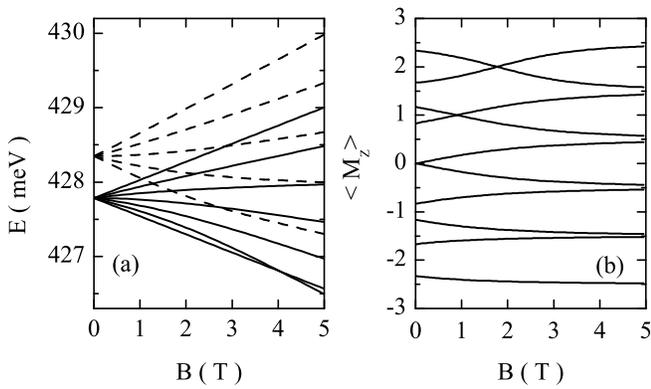}	             
    \caption{ Magnetic field dependence of the: (a) energies of the first twelve energy levels                  
                produced by Zeeman splitting and electron--Mn exchange interaction;
             (b) expected value of the Mn spin along         
             the direction of an applied magnetic field. The system is                  
               considered without SO interaction and                  
              the Mn atom is placed at $\mathbf{R}=\left(7.4, 6.9, 1.5 \right)\:\mathrm{nm}$.  
            }                                                                
    \label{zeeman}                                                           
    \end{figure}

Relevant energies for our problem are the single-particle electron orbital 
energies $E_{i,\sigma}$ [Eq.~(\ref{eq:eis})], the electron--Mn spin 
exchange interaction energy $J_{ij}(\mathbf{R})$ [Eq.~(\ref{eq:jijr})] and
 the Zeeman splitting energy $\mu_{B}B$ [Eq.~(\ref{eq:muBB})]. The separation
between the first two orbital energies is of the order of 50~meV and is 
much larger than the exchange interaction and Zeeman splitting energy which are 
of the order of meV. As a consequence, the first twelve eigenstates of our system
all originate from the ground orbital state and are well separated from higher 
excited states, whose average populations are much smaller even at room temperature. 
The dependence of their energies on magnetic field is shown in 
Fig.~\ref{zeeman}(a). 

Looking at the structure of energy levels for electron--Mn atom system in 
Fig.~\ref{zeeman}(a), several distinctive features can be noticed. Spin of 
an electron is 1/2 and that of a Mn atom is 5/2. Combined, they will give
two possible total spin numbers, \mbox{$F=2$} and \mbox{$F=3$}, respectively with five and 
seven spin projections along the direction of the external magnetic field. When there is no 
magnetic field, total spin is a good quantum number and the presence of 
electron--Mn exchange interaction leads to a splitting 
of the twelve--fold degenerate ground level into two new, seven--fold and 
five-fold degenerate ones. In the case of a finite magnetic field $B$, only $F_z$
remains a good quantum number. The Zeeman term in the Hamiltonian eliminates 
all degeneracies and twelve separate nondegenerate levels emerge. 

A typical state of this system is a superposition of all basis vectors
with the same total spin projection number $F_z$. Because electron--Mn spin coupling
is relatively weak, there are only few dominant states in this linear combination.
 The eigenstate wave
function $|\Psi\rangle$ can be represented as 
    \begin{equation}      
    |\Psi\rangle = \alpha|0, +, M_- \rangle                                  
                 + \beta |0, -, M_+ \rangle +\ldots ,                        
    \label{eigstate}                                                        
    \end{equation}                                                           
where the basis vectors are ordered by the strength of their contribution to the eigenstate. 
In Eq.~(\ref{eigstate}), $|0\rangle$ denotes the ground orbital state, $|+\rangle$
and $|-\rangle$ are the electron spin up and down states, while the Mn spin quantum 
number is related to $F_z$ via $M_{\pm}$=$F_z\pm1/2 \textrm{, where }\ F_z=-2,\ldots,2$. 
%Levels with $F_z=-3$ and $F_z=3$ are not discussed in this part, but will become relevant when 
%SO interaction is included. 
The next term in Eq.~(\ref{eigstate}) depends mostly 
on the Mn--electron coupling strength $J_{0j}(\mathbf{R})$, i.e. on the position of the Mn 
atom.

For the study of Mn (and electron) spin relaxation, the next property that we should turn 
our attention to is the expected value of the Mn spin projection, $M_z$. Its dependence on 
magnetic field for first ten levels with $F_z=-2,\ldots,2$ is presented in Fig.~\ref{zeeman}(b). 
For most values of the magnetic field the expected values of $M_z$ are very close to the corresponding
half integer values from the interval $-5/2$ to $5/2$, suggesting that Mn spin is well defined
for a given eigenstate. This corresponds to the case where one of the 
$\alpha$ and $\beta$ coefficients in Eq.~(\ref{eigstate}) is much larger than the other. In such a case, 
the electron spin projection $S_z=F_z-M_z$  is also well defined.
%Since $F_z$ is exactly a good quantum number and $S_z=F_z-M_z$, while $M_z$ is well defined for 
%most of the states and at most of the magnetic fields, it follows that the same is the case for $S_z$.
$M_z$ and $S_z$ cease to be well defined only for certain fields and for some states where energy 
level crossings occur (see Fig.~\ref{zeeman}). Spin relaxation times calculated within 
our approach should be taken with caution in such cases.

\subsection{Spin relaxation time}\label{Sec:srtsec}

Since electron--phonon interaction Hamiltonian is independent of electron and Mn spin, 
phonons can only induce transitions between energy levels with the
same total spin projection number. Under these conditions, there are only five downhill 
(as well as five uphill) transitions allowed. Each of these downhill transitions 
corresponds to Mn spin--flip from $M_z=F_z+1/2$ to 
$M_z=F_z-1/2$ and in the same time to electron spin--flop from $S_z=-1/2$ to $S_z=1/2$.

The transition time between the two states with the same $F_z$ is therefore 
also the relaxation time for Mn spin--flip from $M_z=F_z+1/2$ to 
$M_z=F_z-1/2$. 
%Since there is only one relevant level with $F_z=-3$ and $F_z=3$
On the other hand, to obtain the electron spin relaxation time, 
one has to take the average over all possible transitions that lead to an electron spin--flip
or spin--flop.

\begin{figure}[bth]                                                      
    \centering                                                               
    \includegraphics[width= 0.49 \textwidth]{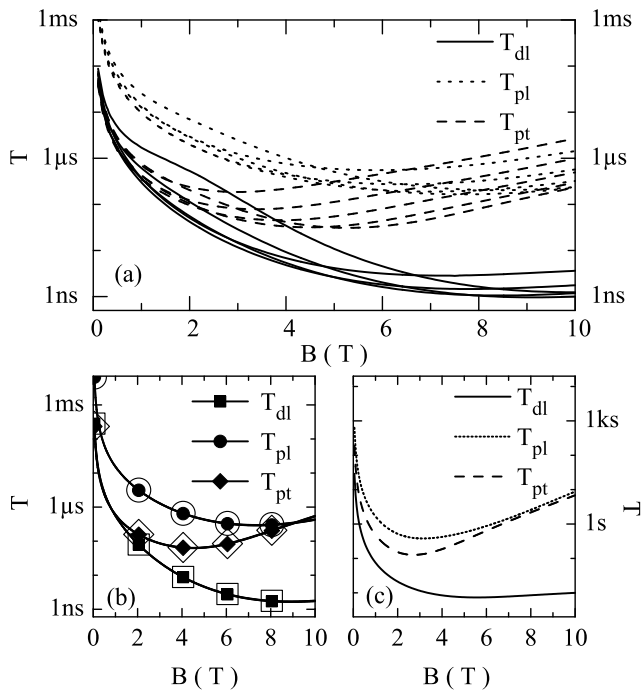}	                     
    \caption{Spin relaxation times for a Mn atom  placed  at $\mathbf{R}=\left(7.4, 6.9, 1.5 \right)\:\mathrm{nm}$
             at room temperature (T=295K), caused by deformation potential interaction
             with LA phonons (dl), piezoelectric potential interaction
             with LA (pl) and TA phonons (pt):                 
            (a) Five transitions corresponding to Mn atom spin-flip.                                         
            (b) Electron spin relaxation time obtained by averaging.     
            Empty symbols represent the data for spin--down    
            to spin--up process, while the filled ones represent 
            the opposite. 
            (c) The same as (b) but for the calculation performed using an inaccurate approximation.}                                           
    \label{srt_no_SO}                                                        
    \end{figure}

The dependence of Mn and electron spin relaxation times on magnetic field in the case of a Mn
atom placed near the center of the dot [at $\mathbf{R}=\left(7.4, 6.9, 1.5 \right)\:\mathrm{nm}$] at room temperature is shown
in Figs.~\ref{srt_no_SO}(a) and \ref{srt_no_SO}(b).

External field affects the degree of Zeeman splitting, which as a consequence 
determines the energy of the phonon through which the system can relax. Since 
a linear dispersion relation connects the phonon energy and its wave
vector, the external field impact on spin relaxation times comes mostly from 
scattering matrix elements [defined in Eqs.~(\ref{eq:sme1})-(\ref{eq:sme3})]. 
For relaxation through  deformation potential, the
scattering matrix element is $\sim q$, while for piezo--field it is $\sim 1/q$. 
Along with the $q^2$ factor that comes from the integration over $\mathbf{q}$ in 
Eq.~(\ref{Eq:eqgif}), this gives an overall $\sim 1/q^3$ dependence for relaxation time 
through deformation potential and $\sim 1/q$ dependence for the relaxation time 
through piezo--field.
As a consequence, spin relaxation becomes more
probable as the external field increases, as can be seen from Fig.~\ref{srt_no_SO}(a).
The above mentioned $\sim 1/q^3$ and $\sim 1/q$ dependences are only approximately followed
because the form factor [Eq.~(\ref{Eq:eqgif})] also depends on $\mathbf{q}$ but this 
dependence is relatively weak in the range of magnetic fields of our interest.
                                                                                                                                                          
Fig.~\ref{srt_no_SO}(a) shows that the results for all possible 
transitions are very similar, ranging from microseconds to nanoseconds when the magnetic 
field varies. This comes from the
fact that phonon energies for each transition are very similar.   
At zero magnetic field, total spin $\mathbf{F}$ (in addition to $F_z$) becomes a good quantum number.
Due to independence of electron--phonon interaction Hamiltonian on spin, the transitions between 
the states with either different $F$ or $F_z$ become forbidden. This leads to infinite relaxation
times at $B=0$ in our model.
                                                                            
Besides the Mn spin, the electron spin relaxation also occurs. 
As mentioned above, a consequence of the spin conserving Hamiltonian 
is the connection between these two. Each time a Mn atom makes a 
\textsl{flip} to a neighboring spin state,
the electron makes a \textsl{flop} and therefore the electron spin relaxation time
 is obtained by averaging over all possible transitions. The similarity 
between these two can be seen from Fig.~\ref{srt_no_SO}(b). 
There are two types of processes for electrons: from spin--down to spin--up state,
and the opposite one. In general there is a difference between the relaxation times for these two but 
only in the case of sufficiently low temperatures.                                

As mentioned earlier, one, or at most two, coefficients in Eq.~(\ref{eigstate}) 
give the dominant contribution to an eigenstate. Both of these coefficients correspond
to the ground orbital state. It is therefore very tempting to introduce an approximation in
which we would reduce the Hilbert space of the system to basis states that originate from orbital 
ground state only. However, this is not
appropriate since such an approximation would yield infinite relaxation times. A less drastic
approximation, where the eigenstates are calculated accurately, but only first two terms in 
Eq.~(\ref{eigstate}) are kept for the calculation of transition rates, is also highly inaccurate. 
As shown in Fig.~\ref{srt_no_SO}(c), it gives unrealistically large spin relaxation times. The results obtained from
 these tempting, but inaccurate approximations, demonstrate the necessity of including a larger number
of orbital states in the calculation and the need for a numerical, rather than analytical, approach 
to the problem.

\subsection{The role of SO interaction}\label{Sec:soint}
The electron (Mn) spin relaxation processes caused by spin-independent electron--phonon interaction are only possible
due to the presence of terms in $\hat{H}_o$ that mix the states of different spin.
Such a term in $\hat{H}_o$ is the electron--Mn exchange interaction $\hat{H}_{m-el}$, which mixes both
the states of different electron spin and the states of different Mn spin. 

Another term which leads to mixing of the states of opposite electron spin is the SO interaction
[Eq.~(\ref{eq:hso})]. The results presented in Sec.~\ref{Sec:srtsec} were obtained while neglecting 
this term and the goal here is to assess the accuracy of such an approximation.

In the presence of SO interaction, $F_z$ is no longer a good quantum number. However, it turns out that the mean value of 
$F_z$ for an eigenstate is very close to the value of $F_z$ in the absence of SO interaction. Therefore, the eigenstates
of the Hamiltonian can still be labeled by the mean value of $F_z$ for a state. 

\begin{figure*}[bt]                                                      
    \centering                                                               
    \includegraphics[width= 0.75 \textwidth]{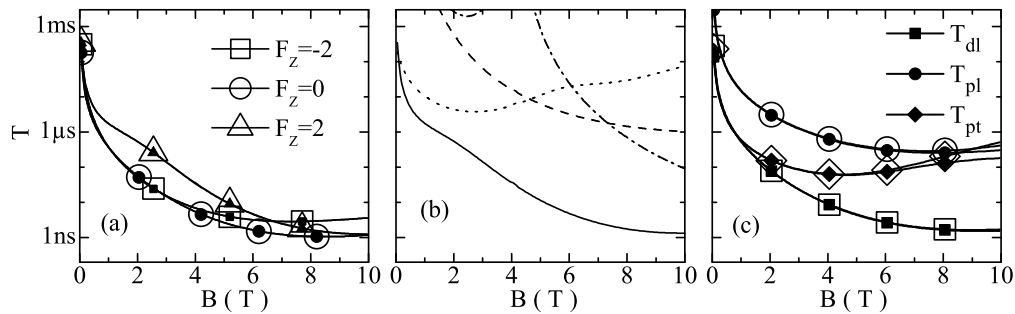}	                     
    \caption{The influence of SO interaction on spin relaxation times for a Mn atom  placed  at $\mathbf{R}=\left(7.4, 6.9, 1.5 \right)\:\mathrm{nm}$.  Magnetic field dependence of relaxation time caused by deformation potential interaction with LA phonons for: (a) the transition between two levels with the same mean value of $F_z$ in the presence (full symbols) and absence (empty symbols) of SO interaction; (b) the transitions between the level ($F=3$, $F_z=2$) and other levels. Full line represents the data for the transition to ($F=2$, $F_z=2$), while dashed, dotted and dash--dotted lines correspond to transitions to ($F=3$, $F_z=1$), ($F=2$, $F_z=1$) and  ($F=3$, $F_z=3$) respectively. The transitions to other levels that are not shown have relaxation times longer than \mbox{$1\:\mathrm{ms}$}. The levels were labeled according to the quantum numbers $F$ and $F_z$ which are good quantum numbers at zero magnetic field in the absence of SO interaction. (c) Magnetic field dependence of average electron spin--up to spin--down relaxation time in the presence (filled symbols) and absence (empty symbols) of SO interaction. The labels dl, pl and pt have the same meaning as in Fig.~\ref{srt_no_SO}. }                                           
    \label{srt_SO}                                                        
    \end{figure*}

In Figs.~\ref{srt_SO}(a) and (c) we compare the electron spin relaxation times in the presence
and absence of SO interaction. The values appear to be rather similar which suggests a weak effect of SO 
interaction on spin relaxation times.

A qualitative difference that SO interaction introduces is that it allows for transitions between the states with different
values of $F_z$, since $F_z$ is no longer a good quantum number. The calculated transition rates for such transitions
from the lowest state with $F_z=2$ to other states are shown in Fig.~\ref{srt_SO}(b). These transitions appear to be much weaker than the transitions
between the states with the same $F_z$.

As a conclusion to this section, we may say that the changes in spin relaxation times due to SO interaction are very small
and for all practical purposes SO interaction can be neglected. Therefore, the rest of the results presented in this paper
will not include SO interaction.

\subsection{Mn position and temperature}\label{sec:postem}

\begin{figure}[bth]                                                      
    \centering                                                               
    \includegraphics[width= 0.49 \textwidth]{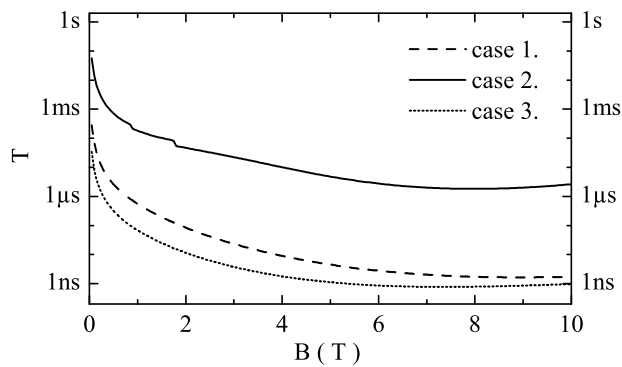}	                     
    \caption{The influence of Mn atom position and temperature on electron spin relaxation times. The dependence of spin relaxation time for the transition from $S_z=1/2$ to $S_z=-1/2$ on magnetic field for different temperatures $T$ and Mn ion positions $\mathbf{R}$ --  
case 1: $T=295$~K, $\mathbf{R}=\left(7.4, 6.9, 1.5 \right)\:\mathrm{nm}$; 
case 2: $T=3$~K,   $\mathbf{R}=\left(7.4, 6.9, 1.5 \right)\:\mathrm{nm}$; 
case 3: $T=295$~K, $\mathbf{R}=\left(4.5, 7.0, 1.5 \right)\:\mathrm{nm}$.
}                                           
    \label{comparison}                                                        
    \end{figure}

The position of Mn atom determines the electron--Mn exchange coupling constants $J_{ij}(\mathbf{R})$ [Eq.~(\ref{eq:jijr})], which is the only place where it appears in the Hamiltonian. Since this is the only term in the Hamiltonian that mixes the states of different electron or Mn spin, one may expect that it has a significant effect on spin relaxation times. To understand the role of the Mn atom position on spin relaxation times, we have performed a calculation for two positions of Mn atom. Position 1 is near the center of the dot at $\mathbf{R}=\left(7.4, 6.9, 1.5\right)\:\mathrm{nm}$, while position 2 was chosen to maximize the $J_{01}(\mathbf{R})$ coupling constant and is given as $\mathbf{R}=\left(4.5, 7, 1.5\right)\:\mathrm{nm}$. The obtained spin relaxation times are shown in Fig.~\ref{comparison}. At position 1, the $J_{01}$, $J_{02}$ and $J_{03}$ coupling constants are nearly zero. As a consequence, the states of interest have the first largest contribution in the linear combination [Eq.~(\ref{eigstate})] other than $\alpha$ and $\beta$ from coefficients corresponding to orbital state $|4\rangle$ [$c^{a}_{4 S_z M_z}$ coefficients in the notation of Eq.~(\ref{eq:psia})]. On the other hand, at position 2, the first largest contribution other than $\alpha$ and $\beta$ comes from $c^{a}_{1 S_z M_z}$. Since state $|1\rangle$ has a lower energy than state $|4\rangle$, the coefficients $c^{a}_{1 S_z M_z}$ for position 2 are larger than $c^{a}_{4 S_z M_z}$ coefficients in the case of position 1. As a consequence, the relevant states at position 2 exhibit a stronger mixing of basis states of different spin, which translates into faster spin relaxation at position 2, as can be seen from Fig.~\ref{comparison}.

Spin relaxation times at temperatures of 3~K and 295~K are also shown in Fig.~\ref{comparison}. When an individual transition is concerned, the temperature appears in our theory only through the phonon occupation number and consequently its effect is easily predictable -- higher temperatures lead to shorter spin relaxation times. When average spin relaxation times are concerned, the temperature appears in the theory also through the thermal weighting factors [Eq.~(\ref{eq:avetau})]. This becomes especially important at higher magnetic fields, where the state with $F_z=3$ becomes the ground state. The transition probability from this state to other states is small but this state has a high weighting factor. The average transition rate is then significantly different than the transition rate for individual transitions between states with the same $F_z$ ($F_z=-2,\ldots,2$). As a consequence of all the mentioned temperature effects, cooling down the system from room temperature to cryogenic temperature leads to an increase in spin relaxation times by three orders of magnitude.

\subsection{Quantum dot dimensions}\label{sec:dim}

\begin{figure}[bth]                                                      
    \centering                                                               
    \includegraphics[width= 0.49 \textwidth]{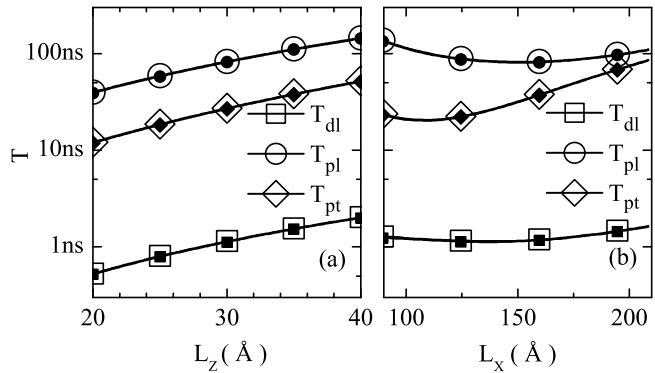}	                     
    \caption{The influence of quantum dot dimensions on electron spin relaxation times at a temperature of $T=295$~K and magnetic field of $B=5$~T. Filled symbols represent the data for spin-up to spin-down process, while the empty ones represent the opposite. The labels dl, pl and pt have the same meaning as in Fig.~\ref{srt_no_SO}. The Mn atom is placed at $\mathbf{R}=\left(4.5, 7.0, 1.5 \right)\:\mathrm{nm}$ for the default dot dimensions and its relative position is kept constant when dot dimensions are varied. (a) The dependence of electron spin relaxation time on the dot dimension in $z$ direction; (b) The dependence of electron spin relaxation time on the dot size in $xy$ plane -- the $L_x/L_y$ ratio and $L_z$ are kept constant, while $L_x$ is varied.}                                           
    \label{fig:fig5}                                                        
    \end{figure}

The dependence of spin relaxation times on quantum dot dimension $L_z$ is shown in Fig.~\ref{fig:fig5}(a).
The relative position of an Mn atom inside the dot is kept during this change of dimensions.
% The fact that the first twelve energy levels have the same dominant electron  orbital state,  leads to a conclusion that the change of the quantum dot size does not produce a drastic change
%in  the  energy  spectrum  structure in the range of energies that is of interest here. 
The electron orbital states [Eq.~(\ref{eq:psieis})] in the range of energies that is of interest here all have quantum number $n_z=1$. As a consequence, their energies do not change when $L_z$ is changed (upto an irrelevant constant).
%and 
%These levels are displaced by approximately equal energy 
%value, and the distance between them will be essentially unchanged. 
%Since the relative position of an Mn atom inside of a dot is kept during this change of dimensions, 
Therefore, the change in $L_z$ affects only the electron wave functions and consequently the electron--Mn exchange coupling constants $J_{ij}(\mathbf{R})$. When the dot dimensions and the confinement volume increases, the probability of finding an electron near the Mn atom, due to normalization of the wave function, decreases. As a consequence, $J_{ij}(\mathbf{R})$ constants decrease. This leads to weaker mixing of states of different spin and therefore spin relaxation times increase, when quantum dot dimensions increase, as can be seen in Fig.~\ref{fig:fig5}(a).

For the same reasons [weaker $J_{ij}(\mathbf{R})$ for larger dots], one may expect that the relaxation times will increase when the dot dimensions in $xy$ plane increase. However, the trend obtained in Fig.~\ref{fig:fig5}(b) is somewhat different. The reason is that when $L_x$ and $L_y$ increase, the distance between electron orbital energies [Eq.~(\ref{eq:eis})] decreases. This leads to stronger contributions from basis states that originate from electron orbitals states other than $|0\rangle$ [the missing terms in Eq.~(\ref{eigstate})]. i.e. to stronger mixing of states of different spin in the expansion from Eq.~(\ref{eigstate}), which leads to shorter spin relaxation times. To summarize, an increase of dot dimensions in $xy$ plane leads on the one hand to a decrease of $J_{ij}(\mathbf{R})$ and on the other hand to an increase in mixing in Eq.~(\ref{eigstate}). As a consequence of these two opposite trends, one obtains a nonmonotonous dependence of spin relaxation times on $L_x$.

\section{Discussion}\label{Sec:discus}

In this section, we discuss the relevance of other possible spin relaxation mechanisms which were not included in our model and compare our results to other relevant results from the literature.

In this work, we have only considered the Dresselhaus SO coupling which is a consequence of bulk inversion asymmetry. Realistic quantum dots also exhibit Rashba SO coupling\cite{pr39-78} as a consequence of structural inversion asymmetry. Our model quantum dot is symmetric, therefore Rashba SO coupling is not present. However, realistic quantum dots certainly exhibit a certain degree of asymmetry and consequently the Rashba SO coupling. The interplay of Dresselhaus and Rashba SO coupling in quantum dots has been studied in Ref.~\onlinecite{prb71-205324}. It was found that both of these are of similar strength and cause similar spin relaxation times. Moreover, in Ref.~\onlinecite{prb61-12639} the authors found that Dresselhaus SO coupling provides a bigger admixture of states of different spin. Since we have found that Dresselhaus SO coupling has practically no effect on spin relaxation rates in our system, it is expected that the same will be the case for Rashba SO coupling.

Spin relaxation by direct electron spin--phonon coupling in GaAs quantum dots was also considered in Refs.~\onlinecite{prb61-12639,prb64-125316}. A conclusion was reached that it is by far less effective than the relaxation caused by SO interaction induced mixing and electron--phonon interaction.

Spin relaxation due to hyperfine interaction between the electron spin and the spin of the nuclei was investigated in Ref.~\onlinecite{prb66-155327}. The hyperfine interaction was found to be the dominant mechanism responsible for spin relaxation rates at magnetic fields which are low enough. Since our model gives relaxation rates that tend to zero at vanishing magnetic field, one may expect that hyperfine interaction will become relevant at low magnetic fields in our system too.

Direct coupling of Mn spin to phonons was considered in Ref.~\onlinecite{prb84-205305} as a potential mechanism of Mn spin relaxation. The corresponding relaxation times  were found to be at least of the order of 1~ms, being much longer than the ones originating from other mechanisms in our work.

Based on the discussion above, we may say that the spin relaxation mechanism considered in our work is certainly the most relevant mechanism in the system studied.

Next, we compare the trends that we obtain for the dependence of relaxation times on various parameters to the ones obtained in quantum dots without Mn atoms.
In Ref.~\onlinecite{prb69-115318}, spin relaxation in GaAs quantum dots caused by the presence of SO interaction and electron--acoustic phonon interaction was investigated. In that situation, spin relaxation takes place due to the transition between the first two Zeeman sublevels of opposite spin. The only spin mixing mechanism in such a system is the SO interaction.         
Spin relaxation times were found to decrease with the increase in magnetic field and were found to be mainly determined by the value of the phonon wave vector responsible for the transition. These conclusions, as one might have expected, are the same in our work. 
%On the other hand, it was found in Ref.~\onlinecite{prb72-115326} that transition times can exhibit maxima at some values of the magnetic field.
Interestingly, it was found on the other hand in Ref.~\onlinecite{prb69-115318} that the dominant electron--phonon interaction mechanism that causes spin relaxation is piezoelectric interaction with TA phonons. In our system, piezoelectric interaction with TA phonons is comparable to deformation potential interaction with LA phonons at low magnetic fields, whereas at higher magnetic fields deformation potential interaction with LA phonons becomes the dominant relaxation mechanism. When the dependence of relaxation times on dot dimensions in $xy$ plane is concerned, it was found in Ref.~\onlinecite{prb69-115318} that these decrease when dot dimensions increase. The presence of exchange interaction between the electron and Mn spin, as the dominant spin mixing mechanism in our system, causes a different trend in our case, as seen in Sec.~\ref{sec:dim}.
It was found that spin relaxation times increase with the increase of dot dimensions in the $z$ direction both in our work and in Ref.~\onlinecite{prb69-115318}, albeit for a different reason. In our case, this is caused by the decrease of electron--Mn spin exchange coupling constant when the dot volume increases (see Sec.~\ref{sec:postem}), while in the work of Ref.~\onlinecite{prb69-115318} this is caused by the decrease of SO interaction.

Finally, we compare the electron spin relaxation times that we obtained to the limits on their values inferred from Mn spin optical orientation experiments.\cite{prb81-245315} In Ref.~\onlinecite{prb82-075321}, a theoretical analysis of Mn spin optical orientation experiments\cite{prb81-245315} was performed to understand the physical origin of the observed Mn spin orientation. It was found that optical orientation that occurs on the $\sim$10ns timescale\cite{prb81-245315} can be explained if hole spin relaxation times of the order of $\sim$10ns are assumed. Similar hole spin relaxation times were obtained from a microscopic theory in Ref.~\onlinecite{prb84-205305}. In the same time, the electron spin relaxation time needs to be longer than that.\cite{prb82-075321} Our theoretical results at a low temperature and low magnetic field ($B<1$~T) (Fig.~\ref{comparison}) indicate electron spin relaxation times longer than tens of microseconds, which is fully consistent with the conclusion obtained in Ref.~\onlinecite{prb82-075321}.

\section{Conclusion}                                   
In conclusion, we found that the interactions responsible for electron and Mn spin relaxation in quantum dots doped with a single Mn atom are the electron--Mn spin exchange interaction and the electron--phonon interaction. The former provides mixing of the states with different spin within an eigenstate and allows for the spin-flip transition caused by the latter. SO interaction has a negligible effect on spin relaxation times, in contrast to conventional quantum dots where SO interaction is the only mechanism which allows for mixing of states of different spin within an eigenstate and the relaxation through the interaction with phonons. Among the different electron--phonon interaction mechanisms, electron interaction with LA phonons turns out to be the dominant mechanism responsible for spin relaxation. Spin relaxation times decrease with the increase in magnetic field. This dependence is mainly determined by an increase in energy level splitting and consequently the increase in the wave vector of the phonon responsible for the transition. The position of Mn ion within the dot determines the strength of electron--Mn spin exchange interaction and therefore significantly alters the spin relaxation times. We find that spin relaxation times in our electron--Mn spin system are longer than in similar hole--Mn spin or exciton--Mn spin systems. This suggests that potential spintronic or quantum computing devices based on the interaction between charge carriers and Mn spin should use electrons as charge carriers.

\section{Acknowledgments}
This work was supported by European Community FP7 Marie Curie Career Integration Grant (ELECTROMAT),
Serbian Ministry of Science (project ON171017) and FP7 projects PRACE-1IP, PRACE-2IP, HP-SEE and 
EGI-InSPIRE.


\begin{thebibliography}{10}%
\makeatletter
\providecommand \@ifxundefined [1]{%
 \ifx #1\undefined \expandafter \@firstoftwo
 \else \expandafter \@secondoftwo
\fi
}%
\providecommand \@ifnum [1]{%
 \ifnum #1\expandafter \@firstoftwo
 \else \expandafter \@secondoftwo
\fi
}%
\providecommand \enquote [1]{``#1''}%
\providecommand \bibnamefont  [1]{#1}%
\providecommand \bibfnamefont [1]{#1}%
\providecommand \citenamefont [1]{#1}%
\providecommand\href[0]{\@sanitize\@href}%
\providecommand\@href[1]{\endgroup\@@startlink{#1}\endgroup\@@href}%
\providecommand\@@href[1]{#1\@@endlink}%
\providecommand \@sanitize [0]{\begingroup\catcode`\&12\catcode`\#12\relax}%
\@ifxundefined \pdfoutput {\@firstoftwo}{%
 \@ifnum{\z@=\pdfoutput}{\@firstoftwo}{\@secondoftwo}%
}{%
 \providecommand\@@startlink[1]{\leavevmode}%
 \providecommand\@@endlink[0]{}%
}{%
 \providecommand\@@startlink[1]{%
  \leavevmode
  \pdfstartlink
   attr{/Border[0 0 1 ]/H/I/C[0 1 1]}%
   user{/Subtype/Link/A<</Type/Action/S/URI/URI(#1)>>}%
  \relax
 }%
 \providecommand\@@endlink[0]{\pdfendlink}%
}%
\providecommand \url  [0]{\begingroup\@sanitize \@url }%
\providecommand \@url [1]{\endgroup\@href {#1}{\urlprefix}}%
\providecommand \urlprefix [0]{URL }%
\providecommand \Eprint[0]{\href }%
\@ifxundefined \urlstyle {%
  \providecommand \doi [1]{doi:\discretionary{}{}{}#1}%
}{%
  \providecommand \doi [0]{doi:\discretionary{}{}{}\begingroup
  \urlstyle{rm}\Url }%
}%
\providecommand \doibase [0]{http://dx.doi.org/}%
\providecommand \Doi[1]{\href{\doibase#1}}%
\providecommand \bibAnnote [3]{%
  \BibitemShut{#1}%
  \begin{quotation}\noindent
    \textsc{Key:}\ #2\\\textsc{Annotation:}\ #3%
  \end{quotation}%
}%
\providecommand \bibAnnoteFile [2]{%
  \IfFileExists{#2}{\bibAnnote {#1} {#2} {\input{#2}}}{}%
}%
\providecommand \typeout [0]{\immediate \write \m@ne }%
\providecommand \selectlanguage [0]{\@gobble}%
\providecommand \bibinfo [0]{\@secondoftwo}%
\providecommand \bibfield [0]{\@secondoftwo}%
\providecommand \translation [1]{[#1]}%
\providecommand \BibitemOpen[0]{}%
\providecommand \bibitemStop [0]{}%
\providecommand \bibitemNoStop [0]{.\EOS\space}%
\providecommand \EOS [0]{\spacefactor3000\relax}%
\providecommand \BibitemShut [1]{\csname bibitem#1\endcsname}%
%</preamble>
\bibitem{rmp76-323}%
  \BibitemOpen
  \bibfield{author}{%
  \bibinfo {author} {\bibfnamefont{I.}~\bibnamefont{\ifmmode \check{Z}\else
  \v{Z}\fi{}uti\ifmmode~\acute{c}\else \'{c}\fi{}}}, \bibinfo {author}
  {\bibfnamefont{J.}~\bibnamefont{Fabian}},\ and\ \bibinfo {author}
  {\bibfnamefont{S.}~\bibnamefont{Das~Sarma}},\ }%
  \bibfield{journal}{%
  \Doi{10.1103/RevModPhys.76.323}{\bibinfo {journal} {Rev. Mod. Phys.}}\ }%
  \textbf{\bibinfo {volume} {76}},\ \bibinfo {pages} {323} (\bibinfo {year}
  {2004})%
  \bibAnnoteFile{NoStop}{rmp76-323}%
\bibitem{pr493-61}%
  \BibitemOpen
  \bibfield{author}{%
  \bibinfo {author} {\bibfnamefont{M.~W.}\ \bibnamefont{Wu}}, \bibinfo {author}
  {\bibfnamefont{J.~H.}\ \bibnamefont{Jiang}},\ and\ \bibinfo {author}
  {\bibfnamefont{M.~Q.}\ \bibnamefont{Weng}},\ }%
  \bibfield{journal}{%
  \bibinfo {journal} {Phys. Rep.}\ }%
  \textbf{\bibinfo {volume} {63}},\ \bibinfo {pages} {61} (\bibinfo {year}
  {2010})%
  \bibAnnoteFile{NoStop}{pr493-61}%
\bibitem{prb84-024402}%
  \BibitemOpen
  \bibfield{author}{%
  \bibinfo {author} {\bibfnamefont{K.~A.}\ \bibnamefont{van Hoogdalem}}\ and\
  \bibinfo {author} {\bibfnamefont{D.}~\bibnamefont{Loss}},\ }%
  \bibfield{journal}{%
  \Doi{10.1103/PhysRevB.84.024402}{\bibinfo {journal} {Phys. Rev. B}}\ }%
  \textbf{\bibinfo {volume} {84}},\ \bibinfo {pages} {024402} (\bibinfo {year}
  {2011})%
  \bibAnnoteFile{NoStop}{prb84-024402}%
\bibitem{Awschalom}%
  \BibitemOpen
  \emph{\bibinfo {title} {Semiconductor Spintronics and Quantum Computation}},\
  edited by\ \bibinfo {editor} {\bibfnamefont{D.~D.}\ \bibnamefont{Awschalom}},
  \bibinfo {editor} {\bibfnamefont{D.}~\bibnamefont{Loss}},\ and\ \bibinfo
  {editor} {\bibfnamefont{N.}~\bibnamefont{Samarth}}\ (\bibinfo {publisher}
  {Springer},\ \bibinfo {year} {2002})%
  \bibAnnoteFile{NoStop}{Awschalom}%
\bibitem{HeuleEtAl}%
  \BibitemOpen
  \bibinfo {note} {R. Heule, C. Bruder, D. Burgarth, and V. M. Stojanovi\'c,
  Phys. Rev. A {\bf 82}, 052333 (2010); Eur. Phys. J. D {\bf 63}, 41 (2011).}%
  \bibAnnoteFile{Stop}{HeuleEtAl}%
\bibitem{Bimberg}%
  \BibitemOpen
  \bibfield{author}{%
  \bibinfo {author} {\bibfnamefont{D.}~\bibnamefont{Bimberg}}, \bibinfo
  {author} {\bibfnamefont{M.}~\bibnamefont{Grundmann}},\ and\ \bibinfo {author}
  {\bibfnamefont{N.~N.}\ \bibnamefont{Ledentsov}},\ }%
  \emph{\bibinfo {title} {Quantum dot heterostructures}}\ (\bibinfo {publisher}
  {John Wiley, Chichester},\ \bibinfo {year} {1999})%
  \bibAnnoteFile{NoStop}{Bimberg}%
\bibitem{Harrison}%
  \BibitemOpen
  \bibfield{author}{%
  \bibinfo {author} {\bibfnamefont{P.}~\bibnamefont{Harrison}},\ }%
  \emph{\bibinfo {title} {{Quantum Wells, Wires and Dots, 2$^{\mathrm{nd}}$
  edition}}}\ (\bibinfo {publisher} {John Wiley and Sons Ltd., Chichester,
  England},\ \bibinfo {year} {2005})%
  \bibAnnoteFile{NoStop}{Harrison}%
\bibitem{prb42-8947}%
  \BibitemOpen
  \bibfield{author}{%
  \bibinfo {author} {\bibfnamefont{U.}~\bibnamefont{Bockelman}}\ and\ \bibinfo
  {author} {\bibfnamefont{G.}~\bibnamefont{Bastard}},\ }%
  \bibfield{journal}{%
  \bibinfo {journal} {Phys. Rev. B}\ }%
  \textbf{\bibinfo {volume} {42}},\ \bibinfo {pages} {8947} (\bibinfo {year}
  {1990})%
  \bibAnnoteFile{NoStop}{prb42-8947}%
\bibitem{prl93-207403}%
  \BibitemOpen
  \bibfield{author}{%
  \bibinfo {author} {\bibfnamefont{L.}~\bibnamefont{Besombes}}, \bibinfo
  {author} {\bibfnamefont{Y.}~\bibnamefont{Leger}}, \bibinfo {author}
  {\bibfnamefont{L.}~\bibnamefont{Maingault}}, \bibinfo {author}
  {\bibfnamefont{D.}~\bibnamefont{Ferrand}}, \bibinfo {author}
  {\bibfnamefont{H.}~\bibnamefont{Mariette}},\ and\ \bibinfo {author}
  {\bibfnamefont{J.}~\bibnamefont{Cibert}},\ }%
  \bibfield{journal}{%
  \bibinfo {journal} {Phys. Rev. Lett.}\ }%
  \textbf{\bibinfo {volume} {93}},\ \bibinfo {eid} {207403} (\bibinfo {year}
  {2004})%
  \bibAnnoteFile{NoStop}{prl93-207403}%
\bibitem{prl95-047403}%
  \BibitemOpen
  \bibfield{author}{%
  \bibinfo {author} {\bibfnamefont{Y.}~\bibnamefont{Leger}}, \bibinfo {author}
  {\bibfnamefont{L.}~\bibnamefont{Besombes}}, \bibinfo {author}
  {\bibfnamefont{L.}~\bibnamefont{Maingault}}, \bibinfo {author}
  {\bibfnamefont{D.}~\bibnamefont{Ferrand}},\ and\ \bibinfo {author}
  {\bibfnamefont{H.}~\bibnamefont{Mariette}},\ }%
  \bibfield{journal}{%
  \bibinfo {journal} {Phys. Rev. Lett.}\ }%
  \textbf{\bibinfo {volume} {95}},\ \bibinfo {eid} {047403} (\bibinfo {year}
  {2005})%
  \bibAnnoteFile{NoStop}{prl95-047403}%
\bibitem{prb71-161307}%
  \BibitemOpen
  \bibfield{author}{%
  \bibinfo {author} {\bibfnamefont{L.}~\bibnamefont{Besombes}}, \bibinfo
  {author} {\bibfnamefont{Y.}~\bibnamefont{Leger}}, \bibinfo {author}
  {\bibfnamefont{L.}~\bibnamefont{Maingault}}, \bibinfo {author}
  {\bibfnamefont{D.}~\bibnamefont{Ferrand}}, \bibinfo {author}
  {\bibfnamefont{H.}~\bibnamefont{Mariette}},\ and\ \bibinfo {author}
  {\bibfnamefont{J.}~\bibnamefont{Cibert}},\ }%
  \bibfield{journal}{%
  \bibinfo {journal} {Phys. Rev. B}\ }%
  \textbf{\bibinfo {volume} {71}},\ \bibinfo {eid} {161307} (\bibinfo {year}
  {2005})%
  \bibAnnoteFile{NoStop}{prb71-161307}%
\bibitem{prb73-045301}%
  \BibitemOpen
  \bibfield{author}{%
  \bibinfo {author} {\bibfnamefont{J.}~\bibnamefont{Fernandez-Rossier}},\ }%
  \bibfield{journal}{%
  \bibinfo {journal} {Phys. Rev. B}\ }%
  \textbf{\bibinfo {volume} {73}},\ \bibinfo {eid} {045301} (\bibinfo {year}
  {2006})%
  \bibAnnoteFile{NoStop}{prb73-045301}%
\bibitem{prb68-045303}%
  \BibitemOpen
  \bibfield{author}{%
  \bibinfo {author} {\bibfnamefont{A.~K.}\ \bibnamefont{Bhattacharjee}}\ and\
  \bibinfo {author} {\bibfnamefont{J.}~\bibnamefont{P\'erez-Conde}},\ }%
  \bibfield{journal}{%
  \Doi{10.1103/PhysRevB.68.045303}{\bibinfo {journal} {Phys. Rev. B}}\ }%
  \textbf{\bibinfo {volume} {68}},\ \bibinfo {pages} {045303} (\bibinfo {year}
  {2003})%
  \bibAnnoteFile{NoStop}{prb68-045303}%
\bibitem{prb71-035338}%
  \BibitemOpen
  \bibfield{author}{%
  \bibinfo {author} {\bibfnamefont{A.~O.}\ \bibnamefont{Govorov}}\ and\
  \bibinfo {author} {\bibfnamefont{A.~V.}\ \bibnamefont{Kalameitsev}},\ }%
  \bibfield{journal}{%
  \bibinfo {journal} {Phys. Rev. B}\ }%
  \textbf{\bibinfo {volume} {71}},\ \bibinfo {eid} {035338} (\bibinfo {year}
  {2005})%
  \bibAnnoteFile{NoStop}{prb71-035338}%
\bibitem{prl102-177403}%
  \BibitemOpen
  \bibfield{author}{%
  \bibinfo {author} {\bibfnamefont{D.~E.}\ \bibnamefont{Reiter}}, \bibinfo
  {author} {\bibfnamefont{T.}~\bibnamefont{Kuhn}},\ and\ \bibinfo {author}
  {\bibfnamefont{V.~M.}\ \bibnamefont{Axt}},\ }%
  \bibfield{journal}{%
  \Doi{10.1103/PhysRevLett.102.177403}{\bibinfo {journal} {Phys. Rev. Lett.}}\
  }%
  \textbf{\bibinfo {volume} {102}},\ \bibinfo {pages} {177403} (\bibinfo {year}
  {2009})%
  \bibAnnoteFile{NoStop}{prl102-177403}%
\bibitem{prl102-127402}%
  \BibitemOpen
  \bibfield{author}{%
  \bibinfo {author} {\bibfnamefont{C.}~\bibnamefont{Le~Gall}}, \bibinfo
  {author} {\bibfnamefont{L.}~\bibnamefont{Besombes}}, \bibinfo {author}
  {\bibfnamefont{H.}~\bibnamefont{Boukari}}, \bibinfo {author}
  {\bibfnamefont{R.}~\bibnamefont{Kolodka}}, \bibinfo {author}
  {\bibfnamefont{J.}~\bibnamefont{Cibert}},\ and\ \bibinfo {author}
  {\bibfnamefont{H.}~\bibnamefont{Mariette}},\ }%
  \bibfield{journal}{%
  \Doi{10.1103/PhysRevLett.102.127402}{\bibinfo {journal} {Phys. Rev. Lett.}}\
  }%
  \textbf{\bibinfo {volume} {102}},\ \bibinfo {pages} {127402} (\bibinfo {year}
  {2009})%
  \bibAnnoteFile{NoStop}{prl102-127402}%
\bibitem{prb81-245315}%
  \BibitemOpen
  \bibfield{author}{%
  \bibinfo {author} {\bibfnamefont{C.}~\bibnamefont{Le~Gall}}, \bibinfo
  {author} {\bibfnamefont{R.~S.}\ \bibnamefont{Kolodka}}, \bibinfo {author}
  {\bibfnamefont{C.~L.}\ \bibnamefont{Cao}}, \bibinfo {author}
  {\bibfnamefont{H.}~\bibnamefont{Boukari}}, \bibinfo {author}
  {\bibfnamefont{H.}~\bibnamefont{Mariette}}, \bibinfo {author}
  {\bibfnamefont{J.}~\bibnamefont{Fern\'andez-Rossier}},\ and\ \bibinfo
  {author} {\bibfnamefont{L.}~\bibnamefont{Besombes}},\ }%
  \bibfield{journal}{%
  \Doi{10.1103/PhysRevB.81.245315}{\bibinfo {journal} {Phys. Rev. B}}\ }%
  \textbf{\bibinfo {volume} {81}},\ \bibinfo {pages} {245315} (\bibinfo {year}
  {2010})%
  \bibAnnoteFile{NoStop}{prb81-245315}%
\bibitem{prl103-087401}%
  \BibitemOpen
  \bibfield{author}{%
  \bibinfo {author} {\bibfnamefont{M.}~\bibnamefont{Goryca}}, \bibinfo {author}
  {\bibfnamefont{T.}~\bibnamefont{Kazimierczuk}}, \bibinfo {author}
  {\bibfnamefont{M.}~\bibnamefont{Nawrocki}}, \bibinfo {author}
  {\bibfnamefont{A.}~\bibnamefont{Golnik}}, \bibinfo {author}
  {\bibfnamefont{J.~A.}\ \bibnamefont{Gaj}}, \bibinfo {author}
  {\bibfnamefont{P.}~\bibnamefont{Kossacki}}, \bibinfo {author}
  {\bibfnamefont{P.}~\bibnamefont{Wojnar}},\ and\ \bibinfo {author}
  {\bibfnamefont{G.}~\bibnamefont{Karczewski}},\ }%
  \bibfield{journal}{%
  \Doi{10.1103/PhysRevLett.103.087401}{\bibinfo {journal} {Phys. Rev. Lett.}}\
  }%
  \textbf{\bibinfo {volume} {103}},\ \bibinfo {pages} {087401} (\bibinfo {year}
  {2009})%
  \bibAnnoteFile{NoStop}{prl103-087401}%
\bibitem{prb82-075321}%
  \BibitemOpen
  \bibfield{author}{%
  \bibinfo {author} {\bibfnamefont{L.}~\bibnamefont{Cywi\ifmmode~\acute{n}\else
  \'{n}\fi{}ski}},\ }%
  \bibfield{journal}{%
  \Doi{10.1103/PhysRevB.82.075321}{\bibinfo {journal} {Phys. Rev. B}}\ }%
  \textbf{\bibinfo {volume} {82}},\ \bibinfo {pages} {075321} (\bibinfo {year}
  {2010})%
  \bibAnnoteFile{NoStop}{prb82-075321}%
\bibitem{prb84-205305}%
  \BibitemOpen
  \bibfield{author}{%
  \bibinfo {author} {\bibfnamefont{C.~L.}\ \bibnamefont{Cao}}, \bibinfo
  {author} {\bibfnamefont{L.}~\bibnamefont{Besombes}},\ and\ \bibinfo {author}
  {\bibfnamefont{J.}~\bibnamefont{Fern\'andez-Rossier}},\ }%
  \bibfield{journal}{%
  \Doi{10.1103/PhysRevB.84.205305}{\bibinfo {journal} {Phys. Rev. B}}\ }%
  \textbf{\bibinfo {volume} {84}},\ \bibinfo {pages} {205305} (\bibinfo {year}
  {2011})%
  \bibAnnoteFile{NoStop}{prb84-205305}%
\bibitem{prb76-245307}%
  \BibitemOpen
  \bibfield{author}{%
  \bibinfo {author} {\bibfnamefont{I.}~\bibnamefont{Savi\ifmmode~\acute{c}\else
  \'{c}\fi{}}}\ and\ \bibinfo {author}
  {\bibfnamefont{N.}~\bibnamefont{Vukmirovi\ifmmode~\acute{c}\else
  \'{c}\fi{}}},\ }%
  \bibfield{journal}{%
  \Doi{10.1103/PhysRevB.76.245307}{\bibinfo {journal} {Phys. Rev. B}}\ }%
  \textbf{\bibinfo {volume} {76}},\ \bibinfo {pages} {245307} (\bibinfo {year}
  {2007})%
  \bibAnnoteFile{NoStop}{prb76-245307}%
\bibitem{crp9-857}%
  \BibitemOpen
  \bibfield{author}{%
  \bibinfo {author} {\bibfnamefont{A.~O.}\ \bibnamefont{Govorov}},\ }%
  \bibfield{journal}{%
  \Doi{10.1016/j.crhy.2008.10.007}{\bibinfo {journal} {C. R. Phys.}}\ }%
  \textbf{\bibinfo {volume} {9}},\ \bibinfo {pages} {857 } (\bibinfo {year}
  {2008})%
  \bibAnnoteFile{NoStop}{crp9-857}%
\bibitem{prl95-217206}%
  \BibitemOpen
  \bibfield{author}{%
  \bibinfo {author} {\bibfnamefont{F.}~\bibnamefont{Qu}}\ and\ \bibinfo
  {author} {\bibfnamefont{P.}~\bibnamefont{Hawrylak}},\ }%
  \bibfield{journal}{%
  \bibinfo {journal} {Phys. Rev. Lett.}\ }%
  \textbf{\bibinfo {volume} {95}},\ \bibinfo {eid} {217206} (\bibinfo {year}
  {2005})%
  \bibAnnoteFile{NoStop}{prl95-217206}%
\bibitem{prb74-245308}%
  \BibitemOpen
  \bibfield{author}{%
  \bibinfo {author} {\bibfnamefont{F.}~\bibnamefont{Qu}}\ and\ \bibinfo
  {author} {\bibfnamefont{P.}~\bibnamefont{Vasilopoulos}},\ }%
  \bibfield{journal}{%
  \Doi{10.1103/PhysRevB.74.245308}{\bibinfo {journal} {Phys. Rev. B}}\ }%
  \textbf{\bibinfo {volume} {74}},\ \bibinfo {pages} {245308} (\bibinfo {year}
  {2006})%
  \bibAnnoteFile{NoStop}{prb74-245308}%
\bibitem{prl96-157201}%
  \BibitemOpen
  \bibfield{author}{%
  \bibinfo {author} {\bibfnamefont{F.}~\bibnamefont{Qu}}\ and\ \bibinfo
  {author} {\bibfnamefont{P.}~\bibnamefont{Hawrylak}},\ }%
  \bibfield{journal}{%
  \bibinfo {journal} {Phys. Rev. Lett.}\ }%
  \textbf{\bibinfo {volume} {96}},\ \bibinfo {eid} {157201} (\bibinfo {year}
  {2006})%
  \bibAnnoteFile{NoStop}{prl96-157201}%
\bibitem{prb72-075358}%
  \BibitemOpen
  \bibfield{author}{%
  \bibinfo {author} {\bibfnamefont{A.~O.}\ \bibnamefont{Govorov}},\ }%
  \bibfield{journal}{%
  \Doi{10.1103/PhysRevB.72.075358}{\bibinfo {journal} {Phys. Rev. B}}\ }%
  \textbf{\bibinfo {volume} {72}},\ \bibinfo {pages} {075358} (\bibinfo {year}
  {2005})%
  \bibAnnoteFile{NoStop}{prb72-075358}%
\bibitem{prb72-075359}%
  \BibitemOpen
  \bibfield{author}{%
  \bibinfo {author} {\bibfnamefont{A.~O.}\ \bibnamefont{Govorov}},\ }%
  \bibfield{journal}{%
  \Doi{10.1103/PhysRevB.72.075359}{\bibinfo {journal} {Phys. Rev. B}}\ }%
  \textbf{\bibinfo {volume} {72}},\ \bibinfo {pages} {075359} (\bibinfo {year}
  {2005})%
  \bibAnnoteFile{NoStop}{prb72-075359}%
\bibitem{prl95-117201}%
  \BibitemOpen
  \bibfield{author}{%
  \bibinfo {author} {\bibfnamefont{J.}~\bibnamefont{Fernandez-Rossier}}\ and\
  \bibinfo {author} {\bibfnamefont{L.}~\bibnamefont{Brey}},\ }%
  \bibfield{journal}{%
  \bibinfo {journal} {Phys. Rev. Lett.}\ }%
  \textbf{\bibinfo {volume} {93}},\ \bibinfo {eid} {117201} (\bibinfo {year}
  {2004})%
  \bibAnnoteFile{NoStop}{prl95-117201}%
\bibitem{prl98-106805}%
  \BibitemOpen
  \bibfield{author}{%
  \bibinfo {author} {\bibfnamefont{J.}~\bibnamefont{Fernandez-Rossier}}\ and\
  \bibinfo {author} {\bibfnamefont{R.}~\bibnamefont{Aguado}},\ }%
  \bibfield{journal}{%
  \bibinfo {journal} {Phys. Rev. Lett.}\ }%
  \textbf{\bibinfo {volume} {98}},\ \bibinfo {eid} {106805} (\bibinfo {year}
  {2007})%
  \bibAnnoteFile{NoStop}{prl98-106805}%
\bibitem{prb78-045321}%
  \BibitemOpen
  \bibfield{author}{%
  \bibinfo {author} {\bibfnamefont{N.~T.~T.}\ \bibnamefont{Nguyen}}\ and\
  \bibinfo {author} {\bibfnamefont{F.~M.}\ \bibnamefont{Peeters}},\ }%
  \bibfield{journal}{%
  \Doi{10.1103/PhysRevB.78.045321}{\bibinfo {journal} {Phys. Rev. B}}\ }%
  \textbf{\bibinfo {volume} {78}},\ \bibinfo {pages} {045321} (\bibinfo {year}
  {2008})%
  \bibAnnoteFile{NoStop}{prb78-045321}%
\bibitem{prb76-045315}%
  \BibitemOpen
  \bibfield{author}{%
  \bibinfo {author} {\bibfnamefont{N.~T.~T.}\ \bibnamefont{Nguyen}}\ and\
  \bibinfo {author} {\bibfnamefont{F.~M.}\ \bibnamefont{Peeters}},\ }%
  \bibfield{journal}{%
  \Doi{10.1103/PhysRevB.76.045315}{\bibinfo {journal} {Phys. Rev. B}}\ }%
  \textbf{\bibinfo {volume} {76}},\ \bibinfo {pages} {045315} (\bibinfo {year}
  {2007})%
  \bibAnnoteFile{NoStop}{prb76-045315}%
\bibitem{prb61-12639}%
  \BibitemOpen
  \bibfield{author}{%
  \bibinfo {author} {\bibfnamefont{A.~V.}\ \bibnamefont{Khaetskii}}\ and\
  \bibinfo {author} {\bibfnamefont{Y.~V.}\ \bibnamefont{Nazarov}},\ }%
  \bibfield{journal}{%
  \Doi{10.1103/PhysRevB.61.12639}{\bibinfo {journal} {Phys. Rev. B}}\ }%
  \textbf{\bibinfo {volume} {61}},\ \bibinfo {pages} {12639} (\bibinfo {year}
  {2000})%
  \bibAnnoteFile{NoStop}{prb61-12639}%
\bibitem{prb66-155327}%
  \BibitemOpen
  \bibfield{author}{%
  \bibinfo {author} {\bibfnamefont{S.~I.}\ \bibnamefont{Erlingsson}}\ and\
  \bibinfo {author} {\bibfnamefont{Y.~V.}\ \bibnamefont{Nazarov}},\ }%
  \bibfield{journal}{%
  \Doi{10.1103/PhysRevB.66.155327}{\bibinfo {journal} {Phys. Rev. B}}\ }%
  \textbf{\bibinfo {volume} {66}},\ \bibinfo {pages} {155327} (\bibinfo {year}
  {2002})%
  \bibAnnoteFile{NoStop}{prb66-155327}%
\bibitem{prb66-161318}%
  \BibitemOpen
  \bibfield{author}{%
  \bibinfo {author} {\bibfnamefont{L.~M.}\ \bibnamefont{Woods}}, \bibinfo
  {author} {\bibfnamefont{T.~L.}\ \bibnamefont{Reinecke}},\ and\ \bibinfo
  {author} {\bibfnamefont{Y.}~\bibnamefont{Lyanda-Geller}},\ }%
  \bibfield{journal}{%
  \Doi{10.1103/PhysRevB.66.161318}{\bibinfo {journal} {Phys. Rev. B}}\ }%
  \textbf{\bibinfo {volume} {66}},\ \bibinfo {pages} {161318} (\bibinfo {year}
  {2002})%
  \bibAnnoteFile{NoStop}{prb66-161318}%
\bibitem{prb67-205330}%
  \BibitemOpen
  \bibfield{author}{%
  \bibinfo {author} {\bibfnamefont{E.}~\bibnamefont{Tsitsishvili}}, \bibinfo
  {author} {\bibfnamefont{R.~v.}\ \bibnamefont{Baltz}},\ and\ \bibinfo {author}
  {\bibfnamefont{H.}~\bibnamefont{Kalt}},\ }%
  \bibfield{journal}{%
  \Doi{10.1103/PhysRevB.67.205330}{\bibinfo {journal} {Phys. Rev. B}}\ }%
  \textbf{\bibinfo {volume} {67}},\ \bibinfo {pages} {205330} (\bibinfo {year}
  {2003})%
  \bibAnnoteFile{NoStop}{prb67-205330}%
\bibitem{prb69-115318}%
  \BibitemOpen
  \bibfield{author}{%
  \bibinfo {author} {\bibfnamefont{J.~L.}\ \bibnamefont{Cheng}}, \bibinfo
  {author} {\bibfnamefont{M.~W.}\ \bibnamefont{Wu}},\ and\ \bibinfo {author}
  {\bibfnamefont{C.}~\bibnamefont{L\"u}},\ }%
  \bibfield{journal}{%
  \Doi{10.1103/PhysRevB.69.115318}{\bibinfo {journal} {Phys. Rev. B}}\ }%
  \textbf{\bibinfo {volume} {69}},\ \bibinfo {pages} {115318} (\bibinfo {year}
  {2004})%
  \bibAnnoteFile{NoStop}{prb69-115318}%
\bibitem{prb69-125330}%
  \BibitemOpen
  \bibfield{author}{%
  \bibinfo {author} {\bibfnamefont{L.~M.}\ \bibnamefont{Woods}}, \bibinfo
  {author} {\bibfnamefont{T.~L.}\ \bibnamefont{Reinecke}},\ and\ \bibinfo
  {author} {\bibfnamefont{R.}~\bibnamefont{Kotlyar}},\ }%
  \bibfield{journal}{%
  \Doi{10.1103/PhysRevB.69.125330}{\bibinfo {journal} {Phys. Rev. B}}\ }%
  \textbf{\bibinfo {volume} {69}},\ \bibinfo {pages} {125330} (\bibinfo {year}
  {2004})%
  \bibAnnoteFile{NoStop}{prb69-125330}%
\bibitem{prb71-075308}%
  \BibitemOpen
  \bibfield{author}{%
  \bibinfo {author} {\bibfnamefont{C.}~\bibnamefont{L\"u}}, \bibinfo {author}
  {\bibfnamefont{J.~L.}\ \bibnamefont{Cheng}},\ and\ \bibinfo {author}
  {\bibfnamefont{M.~W.}\ \bibnamefont{Wu}},\ }%
  \bibfield{journal}{%
  \Doi{10.1103/PhysRevB.71.075308}{\bibinfo {journal} {Phys. Rev. B}}\ }%
  \textbf{\bibinfo {volume} {71}},\ \bibinfo {pages} {075308} (\bibinfo {year}
  {2005})%
  \bibAnnoteFile{NoStop}{prb71-075308}%
\bibitem{prb71-205324}%
  \BibitemOpen
  \bibfield{author}{%
  \bibinfo {author} {\bibfnamefont{D.~V.}\ \bibnamefont{Bulaev}}\ and\ \bibinfo
  {author} {\bibfnamefont{D.}~\bibnamefont{Loss}},\ }%
  \bibfield{journal}{%
  \Doi{10.1103/PhysRevB.71.205324}{\bibinfo {journal} {Phys. Rev. B}}\ }%
  \textbf{\bibinfo {volume} {71}},\ \bibinfo {pages} {205324} (\bibinfo {year}
  {2005})%
  \bibAnnoteFile{NoStop}{prb71-205324}%
\bibitem{prb72-115326}%
  \BibitemOpen
  \bibfield{author}{%
  \bibinfo {author} {\bibfnamefont{C.~F.}\ \bibnamefont{Destefani}}\ and\
  \bibinfo {author} {\bibfnamefont{S.~E.}\ \bibnamefont{Ulloa}},\ }%
  \bibfield{journal}{%
  \Doi{10.1103/PhysRevB.72.115326}{\bibinfo {journal} {Phys. Rev. B}}\ }%
  \textbf{\bibinfo {volume} {72}},\ \bibinfo {pages} {115326} (\bibinfo {year}
  {2005})%
  \bibAnnoteFile{NoStop}{prb72-115326}%
\bibitem{prb72-155333}%
  \BibitemOpen
  \bibfield{author}{%
  \bibinfo {author} {\bibfnamefont{E.}~\bibnamefont{Tsitsishvili}}, \bibinfo
  {author} {\bibfnamefont{R.}~\bibnamefont{v.~Baltz}},\ and\ \bibinfo {author}
  {\bibfnamefont{H.}~\bibnamefont{Kalt}},\ }%
  \bibfield{journal}{%
  \Doi{10.1103/PhysRevB.72.155333}{\bibinfo {journal} {Phys. Rev. B}}\ }%
  \textbf{\bibinfo {volume} {72}},\ \bibinfo {pages} {155333} (\bibinfo {year}
  {2005})%
  \bibAnnoteFile{NoStop}{prb72-155333}%
\bibitem{prb75-041306}%
  \BibitemOpen
  \bibfield{author}{%
  \bibinfo {author} {\bibfnamefont{O.}~\bibnamefont{Olendski}}\ and\ \bibinfo
  {author} {\bibfnamefont{T.~V.}\ \bibnamefont{Shahbazyan}},\ }%
  \bibfield{journal}{%
  \Doi{10.1103/PhysRevB.75.041306}{\bibinfo {journal} {Phys. Rev. B}}\ }%
  \textbf{\bibinfo {volume} {75}},\ \bibinfo {pages} {041306} (\bibinfo {year}
  {2007})%
  \bibAnnoteFile{NoStop}{prb75-041306}%
\bibitem{prb72-075303}%
  \BibitemOpen
  \bibfield{author}{%
  \bibinfo {author} {\bibfnamefont{W.}~\bibnamefont{Yang}}\ and\ \bibinfo
  {author} {\bibfnamefont{K.}~\bibnamefont{Chang}},\ }%
  \bibfield{journal}{%
  \Doi{10.1103/PhysRevB.72.075303}{\bibinfo {journal} {Phys. Rev. B}}\ }%
  \textbf{\bibinfo {volume} {72}},\ \bibinfo {pages} {075303} (\bibinfo {year}
  {2005})%
  \bibAnnoteFile{NoStop}{prb72-075303}%
\bibitem{Mahan}%
  \BibitemOpen
  \bibfield{author}{%
  \bibinfo {author} {\bibfnamefont{G.}~\bibnamefont{Mahan}},\ }%
  \emph{\bibinfo {title} {Many-Particle Physics}}\ (\bibinfo {publisher}
  {Kluwer Academic},\ \bibinfo {year} {2000})%
  \bibAnnoteFile{NoStop}{Mahan}%
\bibitem{pr100-580}%
  \BibitemOpen
  \bibfield{author}{%
  \bibinfo {author} {\bibfnamefont{G.}~\bibnamefont{Dresselhaus}},\ }%
  \bibfield{journal}{%
  \Doi{10.1103/PhysRevB.71.205324}{\bibinfo {journal} {Phys. Rev.}}\ }%
  \textbf{\bibinfo {volume} {100}},\ \bibinfo {pages} {580} (\bibinfo {year}
  {1955})%
  \bibAnnoteFile{NoStop}{pr100-580}%
\bibitem{prb38-1806}%
  \BibitemOpen
  \bibfield{author}{%
  \bibinfo {author} {\bibfnamefont{M.}~\bibnamefont{Cardona}}, \bibinfo
  {author} {\bibfnamefont{N.~E.}\ \bibnamefont{Christensen}},\ and\ \bibinfo
  {author} {\bibfnamefont{G.}~\bibnamefont{Fasol}},\ }%
  \bibfield{journal}{%
  \Doi{10.1103/PhysRevB.38.1806}{\bibinfo {journal} {Phys. Rev. B}}\ }%
  \textbf{\bibinfo {volume} {38}},\ \bibinfo {pages} {1806} (\bibinfo {year}
  {1988})%
  \bibAnnoteFile{NoStop}{prb38-1806}%
\bibitem{bornstein}%
  \BibitemOpen
  \bibinfo {howpublished} {{\textit{Numerical Data and Functional Relationship
  in Science and Technology}, edited by K.~H.~Hellwege, Landolt-B\"ornstein,
  New Series, Group III, Vol. 17, Pt. b (Springer-Verlag, Berlin,1982), p.
  113-149.}}%
  \bibAnnoteFile{Stop}{bornstein}%
\bibitem{prb72-235332}%
  \BibitemOpen
  \bibfield{author}{%
  \bibinfo {author} {\bibfnamefont{S.-J.}\ \bibnamefont{Cheng}},\ }%
  \bibfield{journal}{%
  \Doi{10.1103/PhysRevB.72.235332}{\bibinfo {journal} {Phys. Rev. B}}\ }%
  \textbf{\bibinfo {volume} {72}},\ \bibinfo {eid} {235332} (\bibinfo {year}
  {2005})%
  \bibAnnoteFile{NoStop}{prb72-235332}%
\bibitem{pr39-78}%
  \BibitemOpen
  \bibfield{author}{%
  \bibinfo {author} {\bibfnamefont{Y.~L.}\ \bibnamefont{Bychkov}}\ and\
  \bibinfo {author} {\bibfnamefont{E.~I.}\ \bibnamefont{Rashba}},\ }%
  \bibfield{journal}{%
  \Doi{10.1103/PhysRevB.71.205324}{\bibinfo {journal} {Phys. Rev.}}\ }%
  \textbf{\bibinfo {volume} {39}},\ \bibinfo {pages} {78} (\bibinfo {year}
  {1984})%
  \bibAnnoteFile{NoStop}{pr39-78}%
\bibitem{prb64-125316}%
  \BibitemOpen
  \bibfield{author}{%
  \bibinfo {author} {\bibfnamefont{A.~V.}\ \bibnamefont{Khaetskii}}\ and\
  \bibinfo {author} {\bibfnamefont{Y.~V.}\ \bibnamefont{Nazarov}},\ }%
  \bibfield{journal}{%
  \Doi{10.1103/PhysRevB.64.125316}{\bibinfo {journal} {Phys. Rev. B}}\ }%
  \textbf{\bibinfo {volume} {64}},\ \bibinfo {pages} {125316} (\bibinfo {year}
  {2001})%
  \bibAnnoteFile{NoStop}{prb64-125316}%
\end{thebibliography}
\end{document}